\documentclass[aps]{revtex4-2}
\usepackage{graphicx}
\usepackage{color}
\usepackage{amsmath}
\usepackage{amsfonts}
\usepackage{enumitem}
\usepackage{hyperref}
\hypersetup{
	colorlinks = true,
	citecolor = blue,
	linkcolor= blue
}
\usepackage{color}

\newcommand{\be}{\begin{equation}}
\newcommand{\ee}{\end{equation}}

\newcommand{\bea}{\begin{eqnarray}}
\newcommand{\eea}{\end{eqnarray}}

\renewcommand{\phi}{\varphi}
\renewcommand{\epsilon}{\varepsilon}

\usepackage[normalem]{ulem}
\usepackage{orcidlink}

\begin{document}

\title{Bound states of quasiparticles with quartic dispersion in an external potential: WKB approach}
\date{\today}

\author{E.V. Gorbar\,\orcidlink{0000-0002-2684-1276}}\thanks{Author to whom any correspondence should be addressed}
\affiliation{Department of Physics, Taras Shevchenko National University of Kyiv, 64/13, Volodymyrska Street, Kyiv 01601, Ukraine}
\affiliation{Bogolyubov Institute for Theoretical Physics, 14-b Metrolohichna Street, Kyiv 03143, Ukraine\\E-mail: gorbar@knu.ua}

\author{V.P. Gusynin\,\orcidlink{0000-0003-2378-3821}}
\affiliation{Bogolyubov Institute for Theoretical Physics, 14-b Metrolohichna Street, Kyiv 03143, Ukraine}

\begin{abstract}
The Wentzel-Kramers-Brillouin semiclassical method is formulated for quasiparticles with quartic-in-momentum dispersion which presents the simplest
case of a soft energy-momentum dispersion.  It is shown that matching wave functions in the classically
forbidden and allowed regions requires the consideration of higher-order Airy-type functions. The asymptotics of these functions are
found by using the method of steepest descents  and contain additional exponentially suppressed contributions known as hyperasymptotics.
These hyperasymptotics are crucially important for the correct matching of wave functions in vicinity of turning points for higher-order
differential equations. A quantization condition for bound state energies is obtained, which  generalizes the standard
Bohr-Sommerfeld quantization condition for particles with quadratic energy-momentum dispersion and contains non-perturbative in $\hbar$ correction.
This non-perturbative correction, usually associated with tunneling effects or the presence of complex turning points, occurs even for the harmonic
potential with quartic dispersion where complex turning points and tunneling are absent. The quantization condition is used to find
bound state energies in the case of quadratic and quartic potentials.
\end{abstract}

\maketitle

\section{Introduction}

One of the central topics of modern condensed matter physics is the study of electronic systems  where the potential energy dominates
over the electron kinetic energy. For a given potential energy, this can be realized considering higher order kinetic energy dispersion
$E(p) \sim {p}^{2n}$. Obviously, the kinetic energy flattens at small $p$ as $n$ increases. An interesting example
of the system with such a soft kinetic energy dispersion is provided by ABC-stacked $N$-layer graphene \cite{Guinea2006,Min,Nakamura,Koshino,Cote}
whose low-energy dispersion, neglecting the trigonal warping effects, is given by $E(p)=(v_Fp)^N/\gamma^{N-1}_1$, where $p=\sqrt{p^2_x+p^2_y}$,
$v_F \sim 10^6$ m/s is the Fermi velocity in graphene, and $\gamma_1\approx 0.39$ eV is the nearest interlayer neighbor hopping.

For quasiparticles with soft dispersion $E(p) \sim {p}^{2n}$, where $n \geq 2$, finding solutions of the corresponding higher
order differential equation in the presence of an interaction potential presents a considerable mathematical challenge. The semiclassical Wentzel-Kramers-Brillouin (WKB) approach is an efficient method for determining approximate solutions to linear differential  equations with
spatially varying coefficients \cite{Landau-Lifshitz,Bender-book,Weinberg}. This motivates us to apply this method to determine bound state
energies for
quasiparticles with quartic dispersion, $E(p)\sim p^4$. The cubic dispersion for fermions was also considered in the literature,
e.g., in Refs.\cite{Guinea2006,Heikkila,Liu}. However, the kinetic term for the cubic (or more generally, odd power) dispersion in spaces with
dimension higher than one and rotationally invariant systems has the form $(\mathbf{p}^2)^{3/2}$ which is a non-local (pseudodifferential)
operator in coordinate space. In the 1D case, the third (in general, odd) order differential equation essentially differs from the equation
with four (even) derivatives and requires a separate study.

In the last decades, the WKB method for the one-dimensional Schrödinger equation has showed a significant progress.  One should mention such
achievements as the method of the Maslov canonical operator \cite{Maslov}, the summation of asymptotic series in the Planck constant $\hbar$
(the exact (complex) WKB method \cite{Marino}), the finding of ''instanton-like''  non-perturbative in $\hbar$ corrections to the Bohr-Sommerfeld
quantization condition \cite{Balian,Parisi}. The method has been extended to higher spatial dimensions and multi-component Hamiltonians \cite{Littlejohn}
 and applied to the study of such physical systems as graphene (see, e.g., Refs.\cite{Carmier,Dobrokhotov,Nimyi}).

In this paper, we focus on quasiparticles with quartic dispersion in one space dimension which leads to the study of WKB solutions to a fourth-order
ordinary differential equation with generic potential $V(x)$, where progress in the development and use of the WKB method was not significant. It
should be noted that local WKB solutions of few such equations in low orders in the Planck constant $\hbar$ are available in the literature (see, e.g.,
\cite{Saito1959,Zaslavskiy-book,Fedoryuk-book}), but their global properties, related to quantization, only begin to be investigated \cite{Ito2021}.
Our main interest is a generalization of the Bohr-Sommerfeld quantization condition for bound states which would include exponentially small in $1/\hbar$ corrections.

Since the WKB method is not applicable in vicinity of turning points, to determine energies of bound states by using this method  requires matching
semiclassical wave functions in the classically allowed and forbidden regions and finding the connection formulas. Such a matching technically
is the most complicated part in the realization of the WKB method and proceeds via finding solutions of the equation with linearised potential in
vicinity of turning points with exponentially decreasing asymptotic in the classically forbidden region.

For the quadratic dispersion relation, the corresponding solution is given by the Airy function \cite{Landau-Lifshitz,Weinberg}. In the case of the
quartic dispersion, solutions in vicinity of turning points require a generalization of the Airy functions to solutions of differential equations  of
higher order, which are known as the higher-order Airy functions, or, the hyper-Airy functions. Such functions are given by the Laplace-type integrals
over infinite contours in the complex plane of the integration variable \cite{Ansari2017,Dorugo_thesis}. Not much is known in the literature about these
functions, in particular, their asymptotic behavior in the complex plane. For the problem under consideration, we study the asymptotics on the real axis
as $x\to\pm\infty$ using the extended method of steepest descents \cite{Berry,Paris-Hyperasymptotics} and obtain not only the leading asymptotics,
but also the exponentially suppressed terms (hyperasymptotics) necessary for matching the WKB solutions in vicinity of the turning points. This allows us
to derive a generalization of the Bohr-Sommerfeld quantization condition containing a nonperturbative in $\hbar$ term which affects
most strongly energy values of low-energy states. The found quantization condition is applied to the case of harmonic potential.  By the Fourier
transformation the latter is related to the Schrödinger equation with the quartic potential for which bound state energies are known with great
precision. In addition, the WKB bound state energies are calculated for double quartic system with Hamiltonian $H=a^4p^4+b^4x^4$.

Our approach and obtained results can be extended to the case of kinetic energy dispersion with higher value $n\ge3$. The extension to more space
dimensions is also possible though it is not so straightforward.

The paper is organized as follows. The WKB method is formulated in Sec.\ref{sec:WKB}. The fourth-order Airy functions and their asymptotics using the
steepest descent method are studied in Sec.\ref{sec:generalized}. Matching wave functions via the fourth-order Airy functions, which leads to
a generalization of the Bohr-Sommerfeld quantization condition, is performed in Sec.\ref{sec:matching}. Examples of application of the obtained
quantization condition are considered in Sec.\ref{sec:harmonic-oscillator} for the harmonic oscillator and double quartic systems. The results
are summarized in Sec.\ref{sec:conclusions}. Some useful information about the fourth-order Airy functions, including series and integral representations
as well as their relation to the Wright, Mainardi, and Fax$\rm\acute{e}$n functions, is provided  in Appendix \ref{appendix}.

\section{WKB solutions for quartic dispersion}
\label{sec:WKB}

The Hamiltonian of one-dimensional quasiparticle with quartic dispersion is given by
\begin{align}
H=a^4p^4+V(x),
\label{Hamiltonian}
\end{align}
where $p=-i\hbar \partial_x$ is the momentum operator, $a^4$ is parameter whose dimension is $v^4/W^3$, where $v$ is velocity and $W$ is energy
(cf. with the energy dispersion in ABC-stacked tetra-layer graphene where $E(p)=(v_Fp)^4/\gamma^3_1$). We consider potential $V(x)$ to be a continuous function and $V(x) \to +\infty$ as $|x| \to \infty$.
Therefore, wave functions are three times differentiable. We assume also that $V(x)$ has a single global minimum and ensures the existence of two real turning points. Since Hamiltonian (\ref{Hamiltonian})
is real, wave functions of bound states $\psi$ could be chosen to be real functions. Finally, since we seek to determine bound states, we impose the standard condition that wave functions are square-integrable
which, for potentials  growing as a power function, implies that $\psi(x) \to 0$ for $x \to \pm \infty$.

To solve the stationary Schrödinger-like equation $H\psi=E\psi$ we use the WKB ansatz for the wave function
\begin{align}
\psi(x)=\exp[iS(x)/\hbar],
\label{WKB-ansatz}
\end{align}
and obtain the following equation:
\begin{align}
a^4\left[(S^{\prime})^4-6i\hbar S^{\prime\prime}(S^{\prime})^2-\hbar^2(4S' S'''+3(S'')^2)-i\hbar^3 S^{IV}\right]+V(x)=E,
\label{semiclassical-equation}
\end{align}
where primes denote derivatives with respect to $x$. Expanding $S$ in powers of $\hbar$, $S=\sum_{n=0}^\infty\hbar^n S_n$, one can
get recursive relations for $S_n$. For example, in the zero order, we find the classical local momentum
\begin{equation}
p(x)=S^{\prime}_0(x)=(E-V(x))^{1/4}/a.
\label{semiclassical-momentum}
\end{equation}
For the first and second order corrections, we have the following equations:
\begin{align}
&4(S^{\prime}_0)^3S^{\prime}_1-6iS^{\prime\prime}_0(S^{\prime}_0)^2=0,\\
&6(S_0')^2(S_1')^2+4(S_0')^3S_2'-12i S_0'S_1'S_0''-3(S_0'')^2-6i(S_0')^2S_1''-4S_0'S_0'''=0,
\end{align}
which give
\begin{align}
&S^{\prime}_1=\frac{3iS^{\prime\prime}_0}{2S^{\prime}_0}=\frac{3i}{2}\left(\ln S^{\prime}_0\right)^{\prime}=\frac{3i}{2}\left(\ln p(x)\right)^{\prime},\\
&S^{\prime}_2=\frac{5}{4}\left(\frac{3(p')^2}{2p^3}-\frac{p''}{p^2}\right),
\label{2nd-correction}
\end{align}
where we used $S^{\prime}_0(x)=p(x)$. More high order terms with $n\ge 3$ can be obtained recursively and expressed through $p(x)$ and its derivatives
(for terms with $n\le20$, see Ref.\cite{Ito2021}). In this paper, we restrict our analysis to the zero, first, and second order terms $S_0$, $S_1$, and $S_2$.

Hence, keeping only the terms $S_0$ and $S_1$, the semiclassical wave function equals
\begin{equation}
\psi(x)=\frac{A}{p^{3/2}(x)}e^{\frac{i}{\hbar}\int^x p(u)\,du}, \quad A=const.
\label{semiclassical-wave-function}
\end{equation}

\begin{figure}
\centering
\includegraphics[scale=0.65]{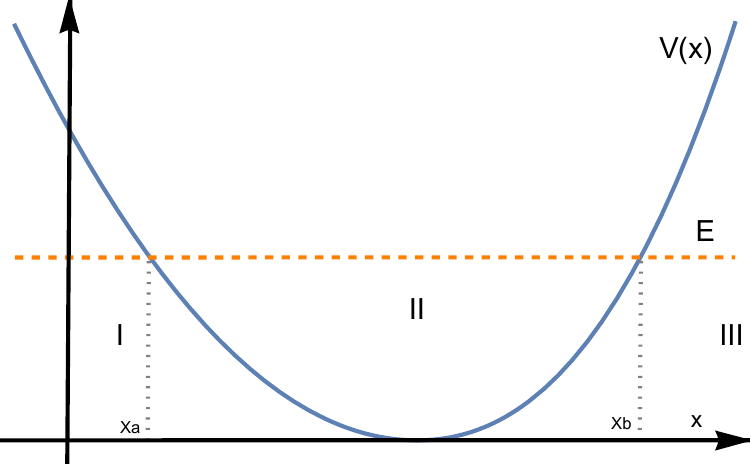}
\caption{Schematics of the bound state problem with potential $V(x)$ (blue solid line), energy $E$ (orange dashed line), turning points $x_a$ and $x_b$, and classically forbidden (\text{I}, \text{III}) and classically allowed (\text{II}) regions.}
\label{fig:potential}
\end{figure}

Let us apply the WKB method to determine bound states for a generic smooth potential with two turning points defined by the equation $E=V(x)$ of the type plotted in
Fig.\ref{fig:potential}. In the classically allowed region
$E \geq V(x)$ with left and right turning points $x_a$ and $x_b$ (we assume that $x_a < x_b$), the classical momentum (\ref{semiclassical-momentum})
equals
\begin{equation}
p_{1,2}(x)=\pm |E-V(x)|^{1/4}/a,\quad p_{3,4}(x)=\pm i|E-V(x)|^{1/4}/a,
\label{momentum-allowed}
\end{equation}
and for each value $p_i$ we have different power series of the function $S(x)$.
Note that the presence of two purely imaginary momenta $p_{3,4}$ in the classically allowed region qualitatively
distinguishes the WKB method for the quartic dispersion from the WKB method for the standard quadratic dispersion.
Clearly, this results in the presence of exponentially increasing and decreasing functions in the general solution for the wave function in
the classically allowed region in addition to the conventional oscillating functions. Since the eigenfunction equation is real, its general
solution can be chosen to be a real function too. We have the following form of the wave function near the turning point $x_a$:
\begin{align}
\psi_a(x)=\frac{1}{p^{3/2}(x)}\left[A_1\cos\left(\frac{1}{\hbar}\int_{x_a}^x p(u)\,du-\frac{\pi}{4}\right)+A_2\sin\left(\frac{1}{\hbar}\int_{x_a}^x p(u)\,du-\frac{\pi}{4}\right)
+A_3e^{\frac{1}{\hbar}\int_{x_a}^x p(u)\,du}+A_4e^{-\frac{1}{\hbar}\int_{x_a}^x p(u)\,du}\right],
\label{second-region-left}
\end{align}
where $A_1$, $A_3$, $A_3$, and $A_4$ are real constants, and $p(u)=|E-V(u)|^{1/4}/a$. In vicinity of the right turning point $x_b$, it is convenient to use another representation of
the wave function in the classically allowed region
\begin{equation}
\psi_b(x)=\frac{1}{p^{3/2}(x)}\left[B_1\cos\left(\frac{1}{\hbar}\int^{x_b}_x p(u)\,du-\frac{\pi}{4}\right)+B_2\sin\left(\frac{1}{\hbar}\int^{x_b}_x p(u)\,du-\frac{\pi}{4}\right)
+B_3e^{\frac{1}{\hbar}\int^{x_b}_x p(u)\,du}+B_4e^{-\frac{1}{\hbar}\int^{x_b}_x p(u)\,du}\right],
\label{second-region}
\end{equation}
where $B_1$, $B_3$, $B_3$, and $B_4$ are real constants.

In the classically forbidden regions $x<x_a$ and $x>x_b$, where $E<V(x)$, the classical momentum (\ref{semiclassical-momentum}) equals
\begin{align}
\frac{1 \pm i}{\sqrt{2}} |E-V(x)|^{1/4}/a,\quad \frac{-1 \pm i}{\sqrt{2}}|E-V(x)|^{1/4}/a
\end{align}
and solutions which decrease at $|x| \to \infty$ are
\begin{equation}
\psi^{(l)}(x)=\frac{e^{-\frac{1}{\hbar\sqrt{2}}\int^{x_a}_x p(u)\,du}}{p^{3/2}(x)}\left(F^{(l)}_1\cos\left(\frac{1}{\hbar\sqrt{2}}\int^{x_a}_x p(u)\,du-\frac{\pi}{8}\right)+F^{(l)}_2\sin\left(\frac{1}{\hbar\sqrt{2}}\int^{x_a}_x p(u)\,du-\frac{\pi}{8}\right)\right),\quad x<x_a
\label{forbidden-left}
\end{equation}
and
\begin{equation}
\psi^{(r)}(x)=\frac{e^{-\frac{1}{\hbar\sqrt{2}}\int^x_{x_b} p(u)\,du}}{p^{3/2}(x)}\left(F^{(r)}_1\cos\left(\frac{1}{\hbar\sqrt{2}}\int^x_{x_b} p(u)\,du-\frac{\pi}{8}\right)+F^{(r)}_2\sin\left(\frac{1}{\hbar\sqrt{2}}\int^x_{x_b} p(u)\,du-\frac{\pi}{8}\right)\right),\quad x>x_b,
\label{forbidden-right}
\end{equation}
respectively. Here $F^{(l)}_1$, $F^{(l)}_2$, $F^{(r)}_1$, and $F^{(r)}_2$ are real constants. In Eqs.(\ref{second-region-left}), (\ref{second-region}) and Eqs.(\ref{forbidden-left}), (\ref{forbidden-right}), we included phases $\pi/4$ and $\pi/8$, respectively, for the convenience of further matching solutions using generalized Airy functions.

Wave functions in the classically allowed and forbidden regions (\ref{second-region})-(\ref{forbidden-right}) should be matched at the turning
points. However, the WKB expansion is not applicable in the vicinity of turning points. Indeed, we can neglect the term of the first order in
$\hbar$ in Eq.(\ref{semiclassical-equation}) only when $|S^{\prime\prime}_0| \ll (S^{\prime}_0 )^2/(6\hbar)$, i.e., for $|(1/p)^{\prime}| \ll 1/(6\hbar)$.
Using the classical momentum (\ref{semiclassical-momentum}), we obtain
\begin{equation}
|V^{\prime}| \ll \frac{|E-V|^{5/4}}{3a\hbar}.
\label{WKB-method-applicability}
\end{equation}
Obviously, this inequality cannot be satisfied at turning points where the classical momentum $p(x)$ vanishes. It is worth comparing  this
inequality with that for validity of the WKB approximation in the case of the quadratic dispersion $E(p)=p^2/(2m)$ where $|V^{\prime}| \ll 2^{3/2}m^{1/2}|E-V|^{3/2}/\hbar$. Near turning points inequality (\ref{WKB-method-applicability}) imposes smaller restriction on $|V^{\prime}|$
compared to the case of the quadratic dispersion because $|E-V|^{5/4}$ is larger than $|E-V|^{3/2}$ as $E-V$ tends to zero.
To illustrate this conclusion, let us consider explicitly the vicinity of a turning point $x_0$ where $V(x) \approx C(x-x_0)$. Then Eq.(\ref{WKB-method-applicability}) gives $(3a)^{4/5}C^{-1/5}\hbar^{4/5} \ll |x-x_0|$ which should be compared with $(8mC)^{-1/3}\hbar^{2/3} \ll |x-x_0|$ in the quadratic dispersion case. Clearly, since the ratio $h^{4/5}/h^{2/3} \to 0$ as $\hbar \to 0$, the WKB approximation is valid closer to the turning points in the quartic dispersion case compared to the quadratic one \cite{range}.

To derive a quantization condition for the bound state energies, it is necessary to match exponentially decreasing solutions in the regions
$x<x_a$ and $x>x_b$ with the solution in the region $x_a<x<x_b$. Two main methods are employed for this purpose. The first is the so-called
Zwaan method \cite{Berry1972,Kemble-book}, where two solutions are matched by analytical continuation of their asymptotics along a path in
the complex $x$ plane going around turning points in the region where the WKB approximation is applicable. The second method allows one to
join solutions, in regions to the left and right of a turning point on the real axis, using a solution of a simpler form of the equation
near the turning point. As is well known, in the case of the quadratic dispersion, matching of wave functions at turning points proceeds via
the Airy function \cite{Landau-Lifshitz,Davydov} whose asymptotes provide the connection formulas. This function is a solution of the Schrödinger equation with $V(x)-E$
replaced near a turning point $x_0$ by $C(x-x_0)$. In the present paper, we use the second method to derive connection formulas for the WKB solutions of the corresponding
fourth-order differential equations.

We would like to comment that to reveal the role of higher order WKB corrections, we include in our analysis in Sec.\ref{sec:harmonic-oscillator} also the second 
order WKB corrections. These corrections, as well as the higher order ones, lead to divergent phase integrals over the classically allowed region. To match solutions in the 
classically allowed and forbidden regions in the presence of such higher-order WKB corrections, Langer’s method of uniformly valid asymptotics in the vicinity of turning points 
could be implemented. For the Schrödinger equation (quadratic dispersion), Langer`s method allows for a fully rigorous treatment of connection formulas avoiding the Zwaan`s 
complex-plane methodology (see, for example, Chapter 4 in \cite{Berry1972}). However, this method is technically  quite complicated for fourth order differential equations and 
is not elaborated yet. Therefore, we will account for the second order WKB corrections by regularizing the corresponding phase integrals and using then an analytical continuation 
procedure.

Finally, before proceeding to the derivation and analysis of bound states in systems with quartic dispersion, it is worth mentioning an interesting 
possibility to realize naturally bound states in the continuum in such systems. For some potentials, like inverted potentials $V(x) \sim- x^2$ or $V(x)\sim -x^4$, the classical momenta $p_{1,2,3,4}$ in Eq.(\ref{momentum-allowed}) describe states in the classically allowed regions which for inverted potentials lie outside the classically forbidden region and extend to infinity. While one of the imaginary solutions (with $p_4$ for positive $x$ and $p_3$ for negative $x$) is not square integrable and should be abandoned, the other is square integrable and potentially can produce a bound state in continuum (the real solutions with $p_1$ and $p_2$ define traveling waves propagating to infinity). In the case of conventional quadratic dispersion, bound states in the continuum exist either for some particular potentials or their existence relies on wave interference for 
suitable chosen parameters which lead to spatially localized states \cite{Hsu,Ahmed}. Since exponentially decaying WKB solutions in the classically allowed region are always present in systems with quartic dispersion, there is no need to fine tune parameters to ensure the square integrability of wave functions in the classically allowed region at spatial infinity. 
 This suggests that bound states in the continuum could be typical and easily realisable in the case of quartic dispersion, e.g., for inverted potentials of general type. This possibility certainly deserves further study.

\section{Fourth-order Airy functions \(Ai_4(x),\,\,\widetilde{Ai}_{4}(x)\) and their asymptotics}
\label{sec:generalized}

\begin{figure}
\centering
\includegraphics[scale=0.54]{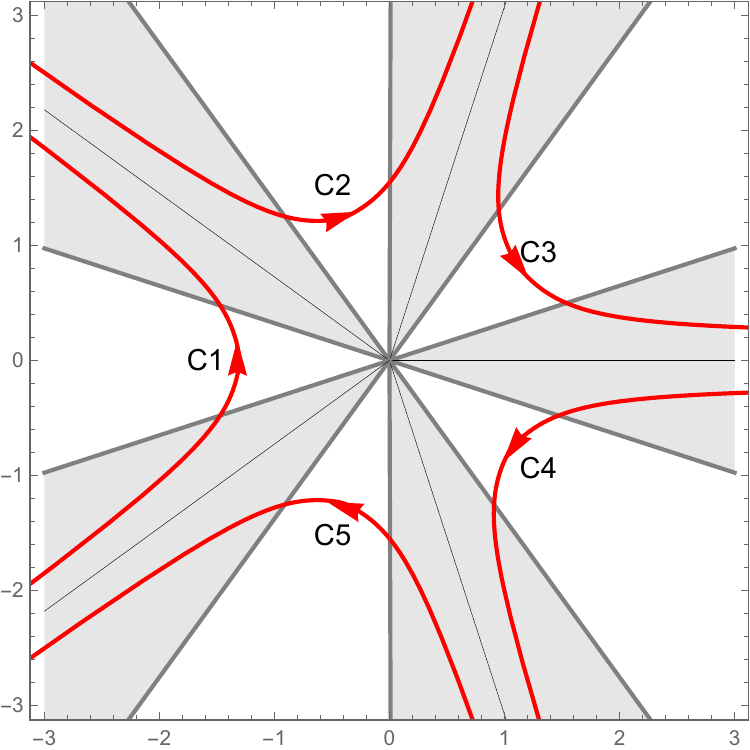}
\caption{The integration paths in the complex plane $t$ defining generalized Airy functions.}
\label{fig:integration-paths}
\end{figure}

To match wave function and determine the connection formulas in our case, we use an approach which utilizes the higher order Airy functions.
Replacing $V-E$ with $C(x-x_b)$ near the right turning point, where $C$ is a positive constant, we find that the wave function in the case of the quartic dispersion satisfies the following equation:
\begin{equation}
a^4\hbar^4\psi^{IV}(x)+C(x-x_b)\psi(x)=0.
\label{quartic-turning-point}
\end{equation}
In terms of variable $z=C^{1/5}(x-x_b)/(\hbar a)^{4/5}$, $C>0$, the above equation takes more simple form
\begin{equation}
\psi^{IV}(z)+z\psi(z)=0.
\label{oscillator-quartic}
\end{equation}
Like in the case of the second order differential equation, solution to the above fourth order differential equation with potential linear in $z$
can be sought in the integral form (Laplace`s method)
\begin{equation}
\psi(z)=const \int_C e^{-zt-\frac{t^5}{5}} dt,
\label{fourth-order-general-solution}
\end{equation}
where infinite contour $C$ in the complex plane $t$ should be chosen so that the integrand tends to zero at the ends of integration contour. This is achieved for $\text{Re}\,t^5 >0$ and, for $t=|t|e^{i\phi}$, this is realised for $\cos(5\phi)>0$, i.e., in 5 sectors listed anticlockwise (see, shaded regions in Fig. \ref{fig:integration-paths}).
%

Since integration contours must begin and end in one of the above sectors, it is convenient to choose five distinct paths $C_j$, $j = 1,\dots 5$, going anticlockwise as shown in Fig.\ref{fig:integration-paths}
\begin{align}
A_j(z)=\frac{1}{2\pi i}\int\limits_{C_j}e^{-zt-\frac{t^5}{5}}dt,\qquad j=\overline{1,5}.
\end{align}
Further, $A_4^*(z)=A_3(z)$, $A_5^*(z)=A_2(z)$ for real $z$, therefore, the function $Ai_4(z)=-A_3(z)-A_4(z)$ is real. This function is an analogue of the well known Airy function. It defines an exponentially decreasing solution for $z \to +\infty$ and can be written in different representations
\begin{align}
Ai_4(z)=-[A_3(z)+A_4(z)]&= -\frac{1}{2\pi i}\int\limits_{C_3+C_4}e^{-zt-\frac{t^5}{5}}dt= \frac{1}{2\pi i}
\int\limits_{e^{-\frac{2\pi}{5}i}\infty}^{e^{\frac{2\pi}{5}i}\infty}e^{-zt-\frac{t^5}{5}}dt\nonumber\\
&=\frac{1}{2\pi i}\int\limits_{-i\infty}^{i\infty}e^{-zt-\frac{t^5}{5}}dt=\frac{1}{\pi}\int\limits_{0}^{\infty}\cos\left(z t+\frac{t^5}{5}\right)dt,
\label{Ai4-integral}
\end{align}
where we deformed the contour $C_3+C_4$ to the imaginary axis in the second line. The function $Ai_4(z)$ is known as the fourth order
Airy function. As to the second real solution, we take (in the notation of Ref.\cite{Dorugo_thesis})
\begin{align}
\widetilde{Ai}_4(z)&=i[A_3(z)-A_4(z)]=\frac{1}{2\pi}\int\limits_{C_3-C_4}e^{-zt-\frac{t^5}{5}}dt=\frac{1}{2\pi }
\left(\int\limits_{e^{\frac{2\pi}{5}i}\infty}^{\infty}\hspace{-2mm}-\hspace{-2mm}\int\limits_{\infty}^{e^{-\frac{2\pi}{5}i}\infty}
\right)e^{-zt-\frac{t^5}{5}}dt\nonumber\\
&=\frac{1}{2\pi }\left(\int\limits_0^\infty + \int\limits_{i\infty}^0-\int\limits_0^{-i\infty}-\int\limits_\infty^0\right)
e^{-zt-\frac{t^5}{5}}dt=\frac{1}{\pi }\int\limits_0^\infty\left[e^{-zt-\frac{t^5}{5}}-\sin\left(zt+\frac{t^5}{5}\right)\right]dt.
\label{repr-Ai4-tilde}
\end{align}

Our choice of functions $Ai_4(z)$ and $\widetilde{Ai}_{4}(z)$ as solutions to Eq.(\ref{fourth-order-general-solution}) is related to their exponentially
decreasing asymptotics in the classically forbidden region of real $ z>0$. As for other two real linearly independent functions, one can
take $Bi_4(z)=-[A_2(z)+A_5(z)]$, $\widetilde{Bi}_4(z)=i[A_2(z)-A_5(z)]$, and these function grow as $z\to+\infty$. These functions are analogous
to the solution $Bi(z)$ for a second-order equation with linear potential. The integral representations for the functions $Ai_4(z)$ and
$\widetilde{Ai}_4(z)$ given above correspond to those in Ref.\cite{Ansari2017} where different notations were used for them.
Alternative integral forms of the solutions can be derived using suitably modified contours of integration \cite{Cinque2024}.

\begin{figure}[ht]
\centering
\includegraphics[scale=0.35]{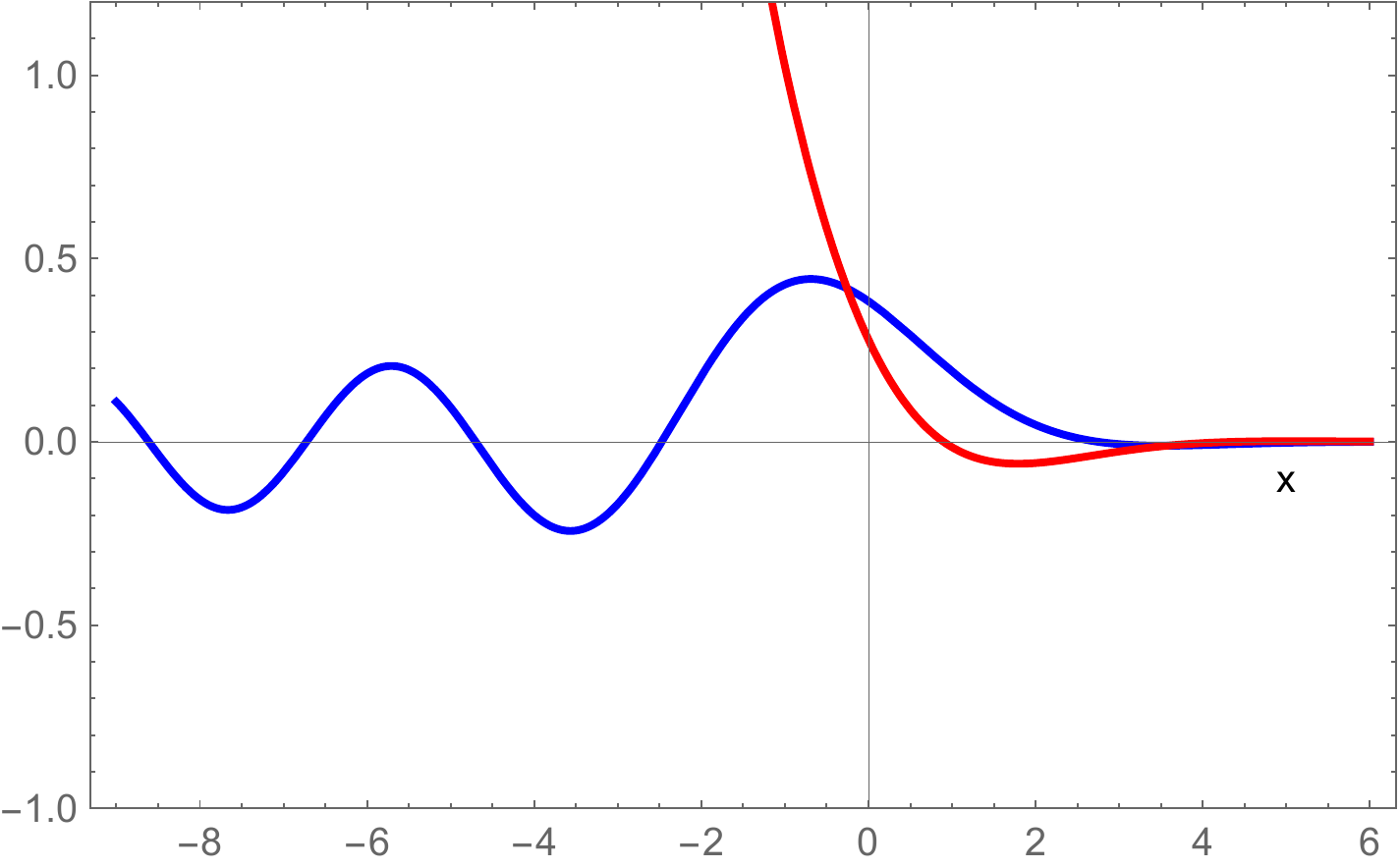}
\caption{The plots of the functions $Ai_4(x)$ (in blue) and $\widetilde{Ai}_4(x)$ (in red) on the real axis.}
\label{fig:Airy_graphs}
\end{figure}
In Fig.\ref{fig:Airy_graphs}, we plot $Ai_4(x)$ and $\widetilde{Ai}_4(x)$ as functions of real argument calculated by using the integral representations
in Eqs.(\ref{Ai4-integral}) and (\ref{repr-Ai4-tilde}) given by integrals along the positive real axis.

\begin{figure}
\centering
\includegraphics[scale=0.33]{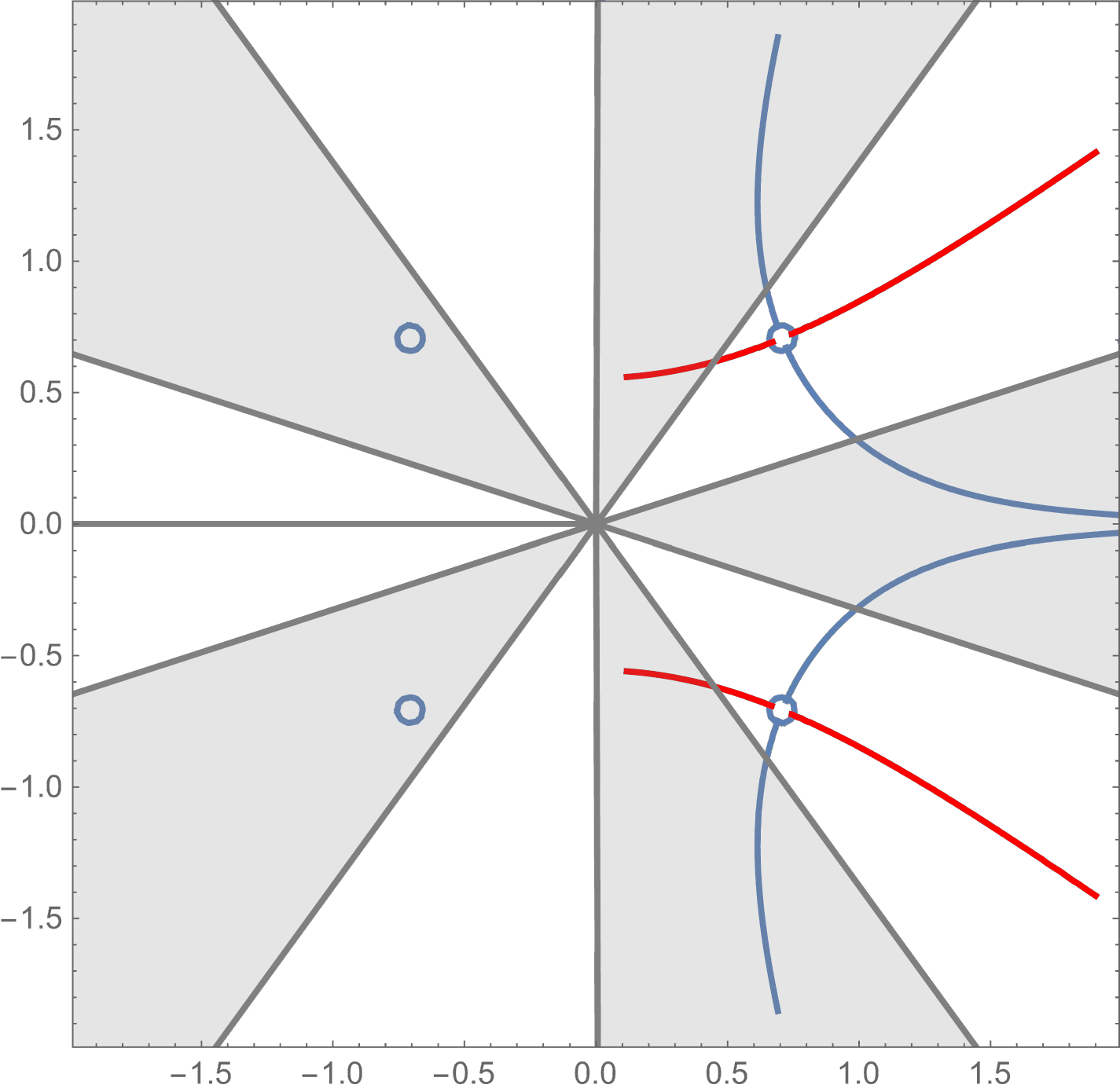}\hspace{5mm}
\includegraphics[scale=0.32]{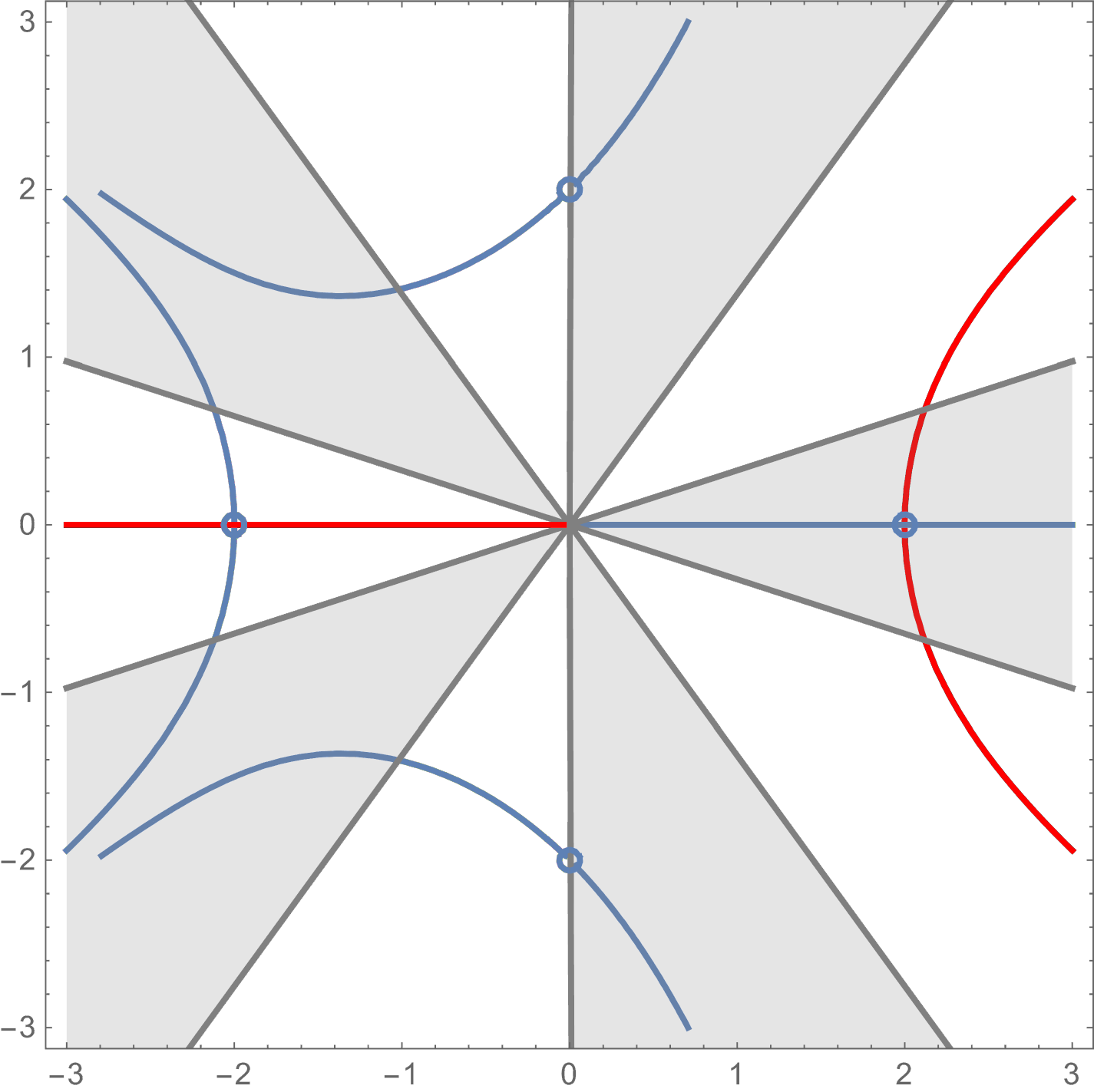}
\caption{Sectors in complex plane $t$ with $\text{Re}\,t^5 >0$ (shaded regions) and extrema of $t+\frac{t^5}{5}$, denoted by small circles, in the classically forbidden region $z>0$ (left panel), and extrema of $t-\frac{t^5}{5}$ in the classically allowed region $z<0$ (right panel). Paths of steepest descent passing through the extrema are shown in blue and paths of strongest ascent in red.}
\label{fig:quartic}
\end{figure}

For $z>0$, we can write the functions $Ai_4(z)$ and $\widetilde{Ai}_4(z)$ in the form
\begin{align}
&Ai_4(z)=-\frac{z^{1/4}}{2\pi i}\int\limits_{C_3+C_4}e^{-z^{5/4}\left( t+\frac{t^5}{5}\right)}dt,\\
&\widetilde{Ai}_4(z)=\frac{z^{1/4}}{2\pi}\int\limits_{C_3-C_4}e^{-z^{5/4}\left( t+\frac{t^5}{5}\right)}dt.
\end{align}
To determine the asymptotics of these functions as $z \to +\infty$ we apply the steepest descent method. For this, we find first
the extrema of the function $t+t^5/5$ in the exponent which are at the points $t_{1,2}=(1\pm i)/\sqrt{2}$, $t_{3,4}=-(1\pm i)/\sqrt{2}$.
The steepest descent path going through the complex conjugate points $t_1$ and $t_2$ is shown in blue in the left panel of Fig.\ref{fig:quartic}.
Obviously, this steepest descent path is a combination of contours $C_3$ and $C_4$ in Fig.\ref{fig:integration-paths}.

As usual in the method of steepest descent, the exponent of the integrand is expanded to the second order in $\delta t=t-t_{sd}$, where $s_{sd}$ is a saddle point and then the resulting Gaussian integral is calculated. Taking into account contributions from the saddle points $t_{1,2}$, we find the asymptotic
\begin{align}
Ai_4(z)\simeq \frac{1}{\sqrt{2\pi}z^{3/8}}e^{-\frac{4z^{5/4}}{5\sqrt{2}}}\cos\left(\frac{4z^{5/4}}{5\sqrt{2}}-\frac{\pi}{8}\right),\qquad z\to +\infty.
\label{Ai4-plus_infty}
\end{align}
As to the function $\widetilde{Ai}_4(z)$, its asymptotic at $z \to +\infty$ is determined by contributions from the same critical points (the steepest descent path shown in blue in Fig.\ref{fig:quartic} is again a combination of contours $C_3$ and $C_4$ in Fig.\ref{fig:integration-paths}).
\begin{eqnarray}
\widetilde{Ai}_4(z)\simeq -\frac{1}{\sqrt{2\pi}\,z^{3/8}}e^{-\frac{4z^{5/4}}{5\sqrt{2}}}\sin\left(\frac{4z^{5/4}}{5\sqrt{2}}-\frac{\pi}{8}\right),\qquad z\to +\infty.
\label{function-right-2}
\end{eqnarray}
Thus, the general solution to equation (\ref{oscillator-quartic}) in the forbidden region $(\text{Re}\,z>0)$ near the right turning point with the correct exponentially decreasing asymptotics is given by
\begin{equation}
\psi^{(l)}_r(z)=C_4Ai_4(z)+\tilde{C}_4\widetilde{Ai}_4(z),
\label{linearized-right-forbidden}
\end{equation}
where $C_4$ and $\tilde{C}_4$ are real constants.

To study the asymptotic of $Ai_4(z)$ and $\widetilde{Ai}_4(z)$ for $z\to-\infty$ we use the representations of these functions in the form
\begin{align}
Ai_4(z)=-\frac{|z|^{1/4}}{2\pi i}\int\limits_{C_3+C_4}e^{|z|^{5/4}\left(t-\frac{t^5}{5}\right)}dt,
\end{align}
\begin{align}
\widetilde{Ai}_4(z)=\frac{|z|^{1/4}}{\pi}\int\limits_0^\infty e^{|z|^{5/4}\left(t-\frac{t^5}{5}\right)}dt-\frac{|z|^{1/4}}{2\pi}\left(\int
\limits_0^{i\infty} e^{|z|^{5/4}\left(t-\frac{t^5}{5}\right)}dt+\int\limits_0^{-i\infty} e^{|z|^{5/4}\left(t-\frac{t^5}{5}\right)}dt\right).
\label{generalized-Airy-representation}
\end{align}

Saddle points are situated now at $t_k=e^{\pi i k/2}$, $k=0,1,2,3$ and the steepest descent paths for the function $f(t)=t-t^5/5$ are determined by the conditions $Re [f(t)-f(t_i)]<0$, $Im [f(t)-f(t_i)]=0$. Setting $t-t_i=\rho\,e^{i\theta_i}$, we find angles at which the steepest descent path traverses the saddle points: $\theta_0=0,\theta_1=\pi/4,\theta_2=\pi/2,\theta_3=-\pi/4$.
Taking into account that the integral over the contour $-C_3-C_4$ equals the integral over the contour $C_1+C_2+C_5$, we deform this contour to the steepest descent path passing through $t_1,t_2,t_3$ (see, the right panel in Fig.\ref{fig:quartic}). Then, according to the steepest descent method, we expand the exponent of the integrand to the second order in deviation from a saddle point and calculate the resulting Gaussian integral. Then taking into account all saddle points contributions, including an exponentially small contribution from the point $t_2=-1$, we find
\begin{align}
Ai_4(z)\simeq\frac{1}{\sqrt{2\pi}(-z)^{3/8}}\cos\left(\frac{4(-z)^{5/4}}{5}-\frac{\pi}{4}\right)+
\frac{e^{-\frac{4(-z)^{5/4}}{5}}}{2\sqrt{2\pi}(-z)^{3/8}},\qquad z\to-\infty.
\label{Ai4-asymp-minusinfinity}
\end{align}

Let us determine now the asymptotics of the function $\widetilde{Ai}_4(z)$ for $z \to -\infty$. In this case three saddle points $t_{0,1,3}=1,i,-i$ lie on the integration contour. Thus, we find the asymptotic
\begin{equation}
\widetilde{Ai}_4(z) \simeq \frac{1}{\sqrt{2\pi}\,(-z)^{3/8}}e^{\frac{4(-z)^{5/4}}{5}}+\frac{1}{\sqrt{2\pi}(-z)^{3/8}}\sin\left(\frac{4(-z)^{5/4}}{5}-\frac{\pi}{4}\right),
\qquad z\to-\infty.
\label{function-left-2}
\end{equation}
The first term is an exponentially increasing function for $z \to -\infty$ and describes the contribution from the saddle point $t_0=1$, while
the second term is due to contributions of $t_1$ and $t_3$. The second terms in Eqs.(\ref{Ai4-asymp-minusinfinity}) and (\ref{function-left-2})
are exponentially suppressed compared to the corresponding first terms and are known as hyperasymptotics \cite{Berry}. The asymptotic behavior of the functions $Ai_4$ and $\widetilde{Ai}_4$ for real argument is also evident in the numerical graphs of these functions in Fig.\ref{fig:Airy_graphs}.

\vskip3mm
\section{Matching wave functions and quantization condition}
\label{sec:matching}

Let us match the solution of the equation with linearised potential (\ref{oscillator-quartic}) found in the previous section to the WKB solutions of the initial equation and determine the quantization condition for bound state energies in the case of a potential with two turning points schematically shown in Fig.\ref{fig:potential} as well as with one turning point and one rigid wall.

\subsubsection{Bound states for two turning points}

We begin matching the wave function near the right turning point given by Eq.(\ref{linearized-right-forbidden}) to the WKB solutions for wave functions in the classically forbidden and allowed regions given by Eqs.(\ref{forbidden-right}) and (\ref{second-region}), respectively. First, we match the wave function (\ref{linearized-right-forbidden}) with the wave function (\ref{forbidden-right}) in the classically forbidden region. For this, we use the asymptotics of the fourth-order Airy functions given in Eqs.(\ref{Ai4-plus_infty}) and (\ref{function-right-2}) and approximate $p(x)=|E-V(x)|^{1/4}/a \approx C^{1/4}(x-x_b)^{1/4}/a$ near the right turning point in the WKB solution (\ref{forbidden-right}). Then we find the relation
\begin{align*}
\frac{1}{\hbar}\int^x_{x_b} p(u)\,du=\frac{4}{5}C^{1/4}(x-x_b)^{5/4}/(\hbar a)=\frac{4}{5}z^{5/4}.
\end{align*}
In the region not far from the turning point $x_b$, where asymptotics (\ref{Ai4-plus_infty}), (\ref{function-right-2}) are still valid, we can equate the functions (\ref{linearized-right-forbidden}) and (\ref{forbidden-right})
which allows us to express constants $F^{(r)}_1$ and $F^{(r)}_2$ of the wave function (\ref{forbidden-right}) through $C_4$ and $\tilde{C}_4$,
\begin{align*}
F^{(r)}_1=\frac{C_4(\hbar aC)^{3/10}}{\sqrt{2\pi}},\quad F^{(r)}_2=-\frac{\tilde{C}_4(\hbar aC)^{3/10}}{\sqrt{2\pi}}.
\end{align*}

Let us match now the wave function (\ref{linearized-right-forbidden}) to the wave function in the classically allowed regions (\ref{second-region}).
Using asymptotics (\ref{Ai4-asymp-minusinfinity}) and (\ref{function-left-2}), we find that the wave function (\ref{linearized-right-forbidden}) in the classically allowed region in vicinity of the right turning point $(x<x_b)$ takes the form
\begin{equation}
\psi^{(l)}_r(z)=\frac{C_4}{(-z)^{3/8}}\cos\left(\frac{4(-z)^{5/4}}{5}
-\frac{\pi}{4}\right)+\frac{C_4}{2(-z)^{3/8}}e^{-\frac{4(-z)^{5/4}}{5}}+\frac{\tilde{C}_4}{(-z)^{3/8}}e^{\frac{4(-z)^{5/4}}{5}}+\frac{\tilde{C}_4}{(-z)^{3/8}}\sin\left(\frac{4(-z)^{5/4}}{5}
-\frac{\pi}{4}\right),
\label{right-turning-point}
\end{equation}
where $z=C^{1/5}(x-x_b)/(\hbar a)^{4/5}$. Further, taking into account that the classical momentum in the classically allowed region near the right turning point $x_b$ equals
$p(x)=|E-V(x)|^{1/4}/a \approx C^{1/4}(x_b-x)^{1/4}/a$ and using the relation
\begin{align*}
\frac{1}{\hbar}\int^{x_b}_x p(u)\,du=\frac{4}{5}C^{1/4}(x_b-x)^{5/4}/(\hbar a)=\frac{4}{5}(-z)^{5/4},
\end{align*}
we obtain that the wave function (\ref{right-turning-point}) can be rewritten in vicinity of the right turning point in the classically allowed region as follows:
\begin{align}
\psi_r(x)=&\frac{(\hbar aC)^{3/10}}{(E-V(x))^{3/8}}\left[C_4\cos\left(\frac{1}{\hbar a}\int^{x_b}_x (E-V(u))^{1/4}du-\frac{\pi}{4}\right)+\frac{C_4}{2}\,e^{-\frac{1}{\hbar a}\int^{x_b}_x (E-V(u))^{1/4}du}\right.\nonumber\\
&+\left.\tilde{C}_4\,e^{\frac{1}{\hbar a}\int^{x_b}_x (E-V(u))^{1/4}du}+\tilde{C}_4\sin\left(\frac{1}{\hbar a}\int^{x_b}_x (E-V(u))^{1/4}du-\frac{\pi}{4}\right)\right].
\label{right-turning-point-1}
\end{align}

Using this wave function, it is not difficult to determine coefficients $B_1$, $B_2$, $B_3$, and $B_4$ of the wave function
$\psi_b(x)$ given by Eq.(\ref{second-region}). We have
\begin{align}
B_1=B_4=\frac{C_4(\hbar aC)^{3/10}}{a^{3/2}},\quad B_2=B_3=\frac{\tilde{C}_4(\hbar aC)^{3/10}}{a^{3/2}}.
\label{coefficients-right}
\end{align}

In vicinity of the left turning point $x_a$, the equation with linearised potential is given by Eq.(\ref{oscillator-quartic}) with $z=C^{1/5}(x_a-x)/(\hbar a)^{4/5}$. The solution of this equation which decreases exponentially in the classically forbidden region as $z\to-\infty$ takes the form
\begin{align*}
\psi^{(l)}_l(z)=D_4Ai_4(-z)+\tilde{D}_4\widetilde{Ai}_4(-z).
\end{align*}
Then repeating similar analysis as done above for the wave function near the right turning point we obtain the following wave function in
the classically allowed region in vicinity of the left turning point:
\begin{align}
\psi_l(x)=&\frac{(\hbar aC)^{3/10}}{(E-V(x))^{3/8}}\left[D_4\cos\left(\frac{1}{\hbar a}\int^x_{x_a} (E-V(u))^{1/4}du-\frac{\pi}{4}\right)+\frac{D_4}{2}\,e^{-\frac{1}{\hbar a}\int^x_{x_a} (E-V(u))^{1/4}du}\right.\nonumber\\
&+\left.\tilde{D}_4\,e^{\frac{1}{\hbar a}\int^x_{x_a} (E-V(u))^{1/4}du}+\tilde{D}_4\sin\left(\frac{1}{\hbar a}\int^x_{x_a} (E-V(u))^{1/4}du-\frac{\pi}{4}\right)\right],
\label{left-turning-point}
\end{align}
which determines coefficients $A_1$, $A_2$, $A_3$, and $A_4$ of the wave function $\psi_a(x)$ given by Eq.(\ref{second-region-left}). We find
\begin{align}
A_1=A_4=\frac{D_4(\hbar aC)^{3/10}}{a^{3/2}},\quad A_2=A_3=\frac{\tilde{D}_4(\hbar aC)^{3/10}}{a^{3/2}}.
\label{coefficients-left}
\end{align}

The wave functions $\psi_r$ and $\psi_l$ given by Eqs.(\ref{right-turning-point-1}) and (\ref{left-turning-point}) and obtained via matching the fourth-order Airy functions near the right and left turning points, respectively, should define the same function in the classically allowed region (equivalently, we could match the wave functions (\ref{second-region-left}) and (\ref{second-region}) with their coefficients defined by Eqs.(\ref{coefficients-right}) and (\ref{coefficients-left})). To perform this matching it is convenient to parametrize the real constants as follows: $C_4=R\sin\alpha$, $\tilde{C}_4=R\cos\alpha$ and $D_4=L\sin\beta$, $\tilde{D}_4=L\cos\beta$. Then using $\int^{x_b}_x=\int^{x_b}_{x_a}-\int^{x}_{x_a}$, we find that oscillating terms in the wave functions $\psi_r(x)$ and $\psi_l(x)$ match when
\begin{align}
R=L,\qquad I=\hbar\left(-\alpha-\beta-\frac{\pi}{2}+2l\pi\right),\quad l=0,\pm 1,\pm 2,...,
\label{matching-oscillating}
\end{align}
where $I=\frac{1}{a}\int^{x_b}_{x_a}(E-V(x))^{1/4}dx$.
Matching exponentially increasing and decreasing terms gives the following equations:
\begin{equation}
\frac{1}{2}\sin\alpha\,\,e^{-I/\hbar}=\cos\beta,\quad \cos\alpha\,\,e^{I/\hbar}=\frac{1}{2}\sin\beta.
\label{matching-exponentials}
\end{equation}
As usual, the overall constant $R$ is determined by the wave function normalization. These equations imply that without loss of generality
we could assume that $\alpha$ and $\beta$ take values on the interval $[0,2\pi]$. The product of two equations in Eq.(\ref{matching-exponentials})
gives $\sin(2\alpha)=\sin(2\beta)$, hence $\beta=\alpha+\pi k$, where $k=0,\pm 1,\pm 2,... $. Since we assume that $\alpha$ and $\beta$ belong to
the interval $[0,2\pi]$, it suffices to restrict $k$ to values $k=0,1$. Further, any of equations in Eq.(\ref{matching-exponentials}) for given $
k$ gives the corresponding value $\alpha$:
\begin{align}
\alpha_k=(-1)^k\arctan\left(2\,e^{I/\hbar}\right).
\end{align}
Then Eq.(\ref{matching-oscillating}) results in the following equation for energy levels:
\begin{align}
I=\hbar\left(-2\alpha_k-\frac{\pi}{2}+2\pi l-\pi k\right).
\label{energy-levels-quadratic}
\end{align}
Finally, combining two series with $l$ and $k=0$ as well as $l$ and $k=1$ into a single series, we obtain the following
quantization condition which defines the bound state energies of the system with quartic dispersion relation:
\begin{align}
\frac{1}{\hbar a}\int^{x_b}_{x_a}(E-V(x))^{1/4}dx=2(-1)^n\arctan\left(\frac{1}{2}e^{-\frac{1}{\hbar a}\int^{x_b}_{x_a}(E-V(x))^{1/4}dx}\right)+\pi(n+\frac{1}{2}),\quad n=0,1,2,...
\label{energy-levels-quadratic-final}
\end{align}
Note that the first term on the right-hand side of the above equation, which is non-perturbative in $\hbar$ and changes sign with $n$, is due to hyperasymptotics
given by the second (exponentially smaller compared to the first) terms in the asymptotics of the fourth-order Airy functions
given by Eqs.(\ref{Ai4-asymp-minusinfinity}) and (\ref{function-left-2}). Higher order perturbative corrections can be taken into account and added to the left-hand side of Eq.(\ref{energy-levels-quadratic-final}). For example, we add the second order correction when we study bound states for the harmonic and quartic potentials in the next section.
\vspace{3mm}

\subsubsection{Bound states for one rigid wall}

It is instructive to determine the quantization condition in the case of one turning point and one rigid wall. The potential in such a system
is taken to be a regular potential $V(x)$ for $x > x_0$ and very large constant potential $V_0$ for $x<x_0$. Using the wave function (\ref{right-turning-point-1}) and matching this function, its first, second, and third derivatives with the exponentially decreasing solution
at $x<x_0$ gives, when taking the limit $V_0 \to +\infty$, the boundary conditions for the wave function $\psi_r(x)$ in the classically allowed
region given by Eq.(\ref{right-turning-point-1})
\begin{equation}
\psi_r(x_0)=0,\quad \psi^{\prime}_r(x_0)=0.
\label{boundary-conditions}
\end{equation}
Thus, we obtain the following equations:
\begin{equation}
C_4\cos\left(\frac{I_w}{\hbar}-\frac{\pi}{4}\right)+\frac{C_4}{2}e^{-I_w/\hbar}
+\tilde{C}_4e^{I_w/\hbar}+\tilde{C}_4\sin\left(\frac{I_w}{\hbar}-\frac{\pi}{4}\right)=0,
\label{rigid-wall-1}
\end{equation}
\begin{equation}
-C_4\sin\left(\frac{I_w}{\hbar}-\frac{\pi}{4}\right)-\frac{C_4}{2}e^{-I_w/\hbar}
+\tilde{C}_4e^{I_w/\hbar}+\tilde{C}_4\cos\left(\frac{I_w}{\hbar}-\frac{\pi}{4}\right)=0,
\label{rigid-wall-2}
\end{equation}
where $I_w=\frac{1}{a}\int^{x_b}_{x_0}(E-V(x))^{1/4}dx$. These equations give the following eigenvalue equation:
\begin{equation}
-\Big[\sin\left(\frac{I_w}{\hbar}-\frac{\pi}{4}\right)-\frac{1}{2}e^{-I_w/\hbar}\Big]\Big[e^{I_w/\hbar}+\sin\left(\frac{I_w}{\hbar}-\frac{\pi}{4}\right)\big]
=\Big[\cos\left(\frac{I_w}{\hbar}-\frac{\pi}{4}\right)
+\frac{1}{2}e^{-I_w/\hbar}\Big]\Big[e^{I_w/\hbar}+\cos\left(\frac{I_w}{\hbar}-\frac{\pi}{4}\right)\Big].
\label{rigid-wall-4}
\end{equation}
This transcendental equation can be easily solved and then we obtain the following quantization condition:
\begin{align}
I_w(E)=\hbar\pi n - \hbar\arcsin\frac{e^{-I_w(E)/\hbar}/\sqrt{2}}{\sqrt{1+e^{-4I_w(E)/\hbar}/4}}-\hbar\arctan\left(e^{-2I_w(E)/\hbar}/2\right).
\end{align}
Keeping only leading in $e^{-I_w(E)/\hbar}$ terms we have
\begin{equation}
\frac{1}{a}\int^{x_b}_{x_0} (E-V(x))^{1/4}dx=\hbar\pi n - \hbar\arcsin\left(\frac{1}{\sqrt{2}}e^{-\frac{1}{\hbar a}\int^{x_b}_{x_0} (E-V(x))^{1/4}dx}\right),\quad n=1,2,3,...\,.
\label{rigid-wall-9}
\end{equation}
Again the presence of the second term, which is non-perturbative in $\hbar$, in the above quantization condition is due
to hyperasymptotics.

\section{Harmonic potential and double quartic system}
\label{sec:harmonic-oscillator}

To check the validity and usefulness of the obtained results, we consider in this section bound states for the quadratic potential
$V(x)=\omega^2x^2/2$ as well as the quartic potential $V(x)=b^4 x^4$. We begin our analysis with the
case of the harmonic potential.

\subsection{Harmonic potential}

The Hamiltonian of a one-dimensional system with quartic dispersion and harmonic potential is given by
$H=a^4p^4+\omega^2x^2/2$.  In terms of dimensionless variable $y=\left(\frac{\omega^2}{2\hbar^4a^4}\right)^{1/6}x$, the eigenvalue equation
takes the simple form
\begin{equation}
(\partial^4_y+y^2)\psi(y)=\varepsilon\psi(y),
\label{equation-harmonic}
\end{equation}
where $\varepsilon=E/(\hbar a\omega/\sqrt{2})^{4/3}$.

\subsubsection{Momentum space representation: quartic oscillator}
\label{sec:momentum}

The harmonic potential is a special case because the eigenvalue equation (\ref{equation-harmonic}) takes the form of the canonical Schrödinger equation with quartic potential in the momentum space representation
\begin{equation}
(-\partial^2_k+k^4)\psi(k)=\varepsilon\psi(k),
\end{equation}
where $k$ is the  wave vector. The WKB analysis of this equation is more involved compared to the case of quadratic potential because
there are four instead of two turning points with two extra turning points situated on the imaginary axis in the complex plane. According
to the analysis in \cite{Balian,Parisi}, these extra turning points generate exponentially small corrections and the energy levels are
defined by the equation (see, for example, Eq.(7) in \cite{Balian})
\begin{equation}
\int_{-\varepsilon^{1/4}}^{\varepsilon^{1/4}}(\varepsilon-k^4)^{1/2}\,dk=\pi(n+\frac{1}{2})+(-1)^n\arctan e^{-\int_{-\varepsilon^{1/4}}^{\varepsilon^{1/4}}(\varepsilon-k^4)^{1/2}\,dk},\quad n=0,1,2,...
\label{momentum-representation}
\end{equation}
If the exponentially small correction is neglected, then the above formula gives the energy levels
\begin{equation}
\varepsilon_n=\left(\frac{(n+\frac{1}{2})\pi}{1.74804}\right)^{4/3}
\label{energy-levels-momentum-representation}
\end{equation}
reproducing the formula from book \cite{Bender-book}. Higher order in $\hbar$ WKB corrections can be added to the left-hand side of Eq.(\ref{momentum-representation}) via the standard procedure.

\subsubsection{Coordinate space representation: quartic energy dispersion}

Applying the quantization condition (\ref{energy-levels-quadratic-final}) obtained in the previous section to the case under consideration given by dimensionless eigenvalue equation (\ref{equation-harmonic}), we find
\begin{equation}
\int^{\sqrt{\varepsilon}}_{-\sqrt{\varepsilon}}(\varepsilon-y^2)^{1/4}dy=\pi(n+\frac{1}{2})+2(-1)^n\arctan\left(\frac{1}{2}e^{-\int^{\sqrt{\varepsilon}}_{-\sqrt{\varepsilon}}(\varepsilon-y^2)^{1/4}dy}\right),\quad n=0,1,2,...
\label{energy-levels-harmonic}
\end{equation}

Obviously, Eqs.(\ref{momentum-representation}) and (\ref{energy-levels-harmonic}) are quite similar in view of the formula
\begin{align*}
\arctan(x)=2\arctan\left(\frac{x}{1+\sqrt{1+x^2}}\right)\approx 2\arctan(\frac{x}{2}), \quad x=e^{-\int^{\sqrt{\varepsilon}}_{-\sqrt{\varepsilon}}(\varepsilon-y^2)^{1/4}dy} \ll 1.
\end{align*}
Eqs.(\ref{momentum-representation}) and (\ref{energy-levels-harmonic}) define the same energy levels if we neglect the $x^2$ term in
the formula above (note that $x$ is indeed quite small) and take into account the equality of the integrals
\begin{align}
\int_{-\varepsilon^{1/4}}^{\varepsilon^{1/4}}(\varepsilon-k^4)^{1/2}\,dk=\int^{\sqrt{\varepsilon}}_{-\sqrt{\varepsilon}}(\varepsilon-y^2)^{1/4}dy=\frac{\sqrt{\pi}\Gamma(1/4)}{3\Gamma(3/4)}\epsilon^{3/4}.
\end{align}
Clearly, this result indicates the validity of our approach which includes the contribution due to hyperasymptotics.
In the case of Schrödinger equation with quartic potential, similar nonperturbative in $\hbar$ contribution appears through
a tunneling effect between complex turning points generating exponentially small corrections \cite{Balian,Parisi}.

It is instructive to present numerical results for semiclassical energy levels in the WKB approximation in two cases: i) neglecting the contribution
due to hyperasymptotics and ii) taking this contribution into account, and then compare the obtained results with the exact numerical values for quartic oscillator given in
\cite{Reid,Banerjee1978}. To elucidate the role of the nonperturbative correction for numerical computations of low-energy levels we compare its contribution with that
due to the second order in $\hbar$ correction for the harmonic and quartic potentials given by Eq.(\ref{2nd-correction}). For the harmonic potential, the second order
correction (\ref{2nd-correction}) becomes
\begin{align}
S'_2(x)=\frac{5(5x^2+4\epsilon)}{32(\epsilon-x^2)^{9/4}}.
\end{align}
The integral of $S'_2(x)$ between the turning points $-\epsilon^{1/2}$ and $\epsilon^{1/2}$ is divergent at the lower and upper limits of integration and requires regularization. The same is true for higher-order corrections $S'_{2n}$, $n\ge2$.

The presence of these divergences means that computing higher-order corrections in the Bohr-Sommerfeld quantization condition requires not only the correction terms in
the expansion $S(x) = S_0(x) + \hbar S_1(x) +...$ but also higher-order analysis of the matching procedure in the vicinity of the turning points. Matching must then be carried out to higher order. Based on experience with other problems (for the quadratic dispersion, see Sec.10.7 in \cite{Bender-book}), we expect that this is equivalent to regularising divergent integrals in higher-order corrections  and using a specific analytical continuation procedure (to be explained below). We believe that this procedure gives the correct higher-order corrections to the Bohr-Sommerfeld quantization condition, and we provide numerical evidence that this is indeed the case.

As mentioned above, to proceed we use the analytic regularization introducing parameter $s$ which makes the integral convergent in some region $\text{Re}\,s>s_0$. Then we calculate the integral and continue analytically the result to the required value of $s$. In the case under consideration, we proceed as follows:
\begin{align}
S_2=\lim_{s\to-2}\frac{5}{32}\int_{-\epsilon^{1/2}}^{\epsilon^{1/2}}(5x^2+4\epsilon)(\epsilon-x^2)^{-1/4+s}dx=
-\frac{\sqrt{\pi}\Gamma(3/4)}{4\Gamma(1/4)\epsilon^{3/4}}.
\end{align}
We check that this result coincides with that obtained by using another analytical continuation, where the integral along the real segment is transformed into an integral over a closed contour in the complex plane, encompassing the turning points, similar to the contour integral used for the exact WKB quantization (see, e.g., Eq.(10) in \cite{Bender}). Thus, we obtain the following transcendental equation for computing energy levels (cf. with Eq.(\ref{energy-levels-harmonic})):
\begin{align}
\frac{\sqrt{\pi}\Gamma(1/4)}{3\Gamma(3/4)}\epsilon^{3/4}-\frac{\sqrt{\pi}\Gamma(3/4)}{4\Gamma(1/4)\epsilon^{3/4}}=\pi\left(n+\frac{1}{2}\right)
+2(-1)^n\arctan\left(\frac{1}{2}\exp\left(-\frac{\sqrt{\pi}\Gamma(1/4)}{3\Gamma(3/4)}\epsilon^{3/4}\right)\right).
\label{double-quartic-energies}
\end{align}

The results for the states with $n=0,1,...,6$  are present in Table I. To quantify the role of non-perturbative contribution, we provide first the bound state energy for the leading-order WKB perturbative contribution $\varepsilon_{lead}$ in the second row and the hyperasymptotics contribution for this case is included in the third row. The second order perturbative correction is taken into account in the 4th row and the 5th row shows $\varepsilon_{2ndhyper}$ where the contribution due to hyperasymptotics is accounted for also. The exact values found in \cite{Reid,Banerjee1978} are given in the last row.

Notice that although Eq.(\ref{double-quartic-energies}) is written in the dimensionless form and does not contain explicitly the Planck constant, the WKB corrections form a series in inverse fractional powers of dimensionless energy $\epsilon$ reflecting the standard WKB expansion in $\hbar$ (recall the definition of $\epsilon$ below Eq.(\ref{equation-harmonic})). Therefore, according this equation, the accuracy of obtained energies increases with the growth of $\epsilon$ (i.e., the level index $n$ should be large enough). Remarkably, our approximation seems to be quite accurate even for $n$ not at all large. This can be seen by comparing the approximate values of energy $\varepsilon$ for the first seven states computed using Eq.(\ref{double-quartic-energies}) with the exact values given in the last row of Table~\ref{table-of-energies-1}.

Comparing these numerical values, we conclude that taking into account  hyperasymptotics notably improves the estimate of the lowest bound state energy. The energy levels $\varepsilon_{lead}$ given by Eq.(\ref{energy-levels-momentum-representation}), which take into account only the leading asymptotics, and the energy levels $\varepsilon_{hyper}$ given by Eq.(\ref{energy-levels-quadratic-final}), which include hyperasymptotics, are practically the same for higher levels and quite close to exact values. Note also that the correction due to hyperasymptotics in the Bohr-Sommerfeld quantization condition is always smaller than $\pi/2$ in accordance with \cite{Turbiner}.

It is clear that the second order WKB correction is important too even for higher energy levels where this correction makes the corresponding numerical values to match practically ideally the exact values of bound state energies.

\begin{table}[ht]
\begin{tabular}{|l|l|l|l|l|l|l|l|}
\hline
$\varepsilon$ \textbackslash \,\,$n$ & 0 & 1 & 2 & 3 & 4 & 5 & 6 \\
\hline
$\varepsilon_{lead}$ & 0.8671  & 3.7519 & 7.4140 & 11.6115 & 16.2336 & 21.2137 & 26.5063  \\
\hline
$\varepsilon_{hyper}$ & 0.9977 & 3.7423 & 7.4145 & 11.6115 & 16.2336 & 21.2137 & 26.5063  \\
\hline
$\varepsilon_{2nd}$ &0.9808 & 3.8103 & 7.4558 & 11.6450 & 16.2619 & 21.2384 & 26.5285\\
\hline
$\varepsilon_{2ndhyper}$ & 1.0907 & 3.8013 & 7.4563 & 11.6450 & 16.2619 & 21.2384 & 26.5285\\
\hline
$\varepsilon_{exact}$ & 1.0604  & 3.7997 & 7.4557 & 11.6447 & 16.2618 & 21.2384 & 26.5285 \\
\hline
\end{tabular}
\caption{Numerical values of the bound state energy $\varepsilon$ for one-dimensional system with quartic dispersion and quadratic
potential for $n=0,1,\dots,6$. The leading-order WKB contribution $\varepsilon_{lead}$ and in addition the hyperasymptotics contribution $\varepsilon_{hyper}$ are shown in the second and third rows, respectively. The numerical value of the bound state energy $\varepsilon_{2nd}$, where the second order WKB correction is included, is given in the 4th row. The 5th row shows $\varepsilon_{2ndhyper}$ where the contribution due to hyperasymptotics is accounted for also. The exact values found in \cite{Reid,Banerjee1978} are given in the last row.}
\label{table-of-energies-1}
\end{table}

\subsection{Double quartic system}
\label{sec:quartic-potential}

The Hamiltonian of a one-dimensional system with quartic dispersion and quartic potential is given by $H=a^4p^4+b^4x^4$. In terms
of dimensionless variable $y=x\sqrt{b/(a\hbar)}$, the eigenvalue equation takes the simple form
\begin{equation}
(\partial^4_y+y^4)\psi(y)=\varepsilon\psi(y),
\label{double-quartic equation}
\end{equation}
where $\varepsilon=E/(\hbar^2a^2b^2)$. In this case, taking into account the leading and second order WKB contributions and the non-perturbative contribution, the energy levels of the double quartic system are defined by the equation
\begin{align}
\frac{\Gamma^2(1/4)\sqrt{\epsilon}}{4\sqrt{\pi}}-\frac{3\Gamma^2(3/4)}{4\sqrt{\pi}}\frac{1}{\sqrt{\epsilon}}=\pi\left(n+\frac{1}{2}\right)+2(-1)^n\arctan\left(\frac{1}{2}\exp\left[-\frac{\Gamma^2(1/4)\sqrt{\epsilon}}{4\sqrt{\pi}}\right]\right),
\qquad n=0,1,\dots\,.
\end{align}

The numerical values of bound state energies with the leading and second order WKB contributions are shown in the second row of Table II. The energy levels $\varepsilon_{2ndhyper}$ which include the non-perturbative contribution are given in the third row. Note that as in the case of the harmonic potential, our asymptotic method requires the quantum number $n$ to be large enough. Similar to the case of the harmonic potential, the difference between keeping only the second-order correction or keeping also the non-perturbative contribution is only notable for the lowest values of $n$. As to the exact energy eigenvalues of the problem under consideration, to the best of our knowledge they are absent in the literature and, therefore, we do not provide them in Table II. For the ground state with $n=0$, we find $\varepsilon=1.4241$ which should be compared with the value $1.3967$ obtained via direct numerical calculation \cite{Grinyuk}.

\begin{table}[ht]
\begin{tabular}{|l|l|l|l|l|l|l|l|}
\hline
$\varepsilon$ \textbackslash \,\,$n$ & 0 & 1 & 2 & 3 & 4 & 5 & 6 \\
\hline
$\varepsilon_{2nd}$ & 1.3138  & 7.1289 & 18.6234 & 35.8529 & 58.8228 & 87.5342 & 121.988  \\
\hline
$\varepsilon_{2ndhyper}$ & 1.4241 & 7.1093 & 18.6249 & 35.8528 & 58.8228 & 87.5342 & 121.988  \\
\hline
\end{tabular}
\caption{Numerical values of the bound state energies $\varepsilon$ for the one-dimensional system with quartic dispersion and quartic potential for $n=0,1,\dots,6$ for the leading and second order WKB contributions $\varepsilon_{2nd}$ and taking into account also the non-perturbative contribution $\varepsilon_{hyper}$.}
\label{table-of-energies-3}
\end{table}

Finally, before drawing our conclusions, it is worth adding that although we considered here explicitly only the cases of harmonic and quartic potentials, bound state energies for general powerlike potential $V(x)=x^{2m}$, $m=1,2,3,..$ and higher-order dispersion $E \sim p^{2n}$ can be easily found. Their energy levels are defined by Eq.(\ref{energy-levels-quadratic-final}) where the integral on the left-hand side should be in dimensionless variables replaced with
\begin{equation}
\int^{\varepsilon^{1/(2m)}}_{-\varepsilon^{1/(2m)}}(\varepsilon-x^{2m})^{1/(2n)}dx=\frac{B(\frac{1}{2m},\frac{1}{2n})}{m+n}\varepsilon^{\frac{m+n}{2mn}}
\label{integral}
\end{equation}
with $B(u,v)$ being the beta-function. As to the non-perturbative contribution due to hyperasymptotics, it should be determined via the analysis of asymptotics of the corresponding higher-order Airy functions.

\section{Summary}
\label{sec:conclusions}

We formulated the semiclassical WKB approach to determine bound states energies for quasiparticles with quartic energy-momentum
dispersion. We determined semiclassical wave functions both in the
classically allowed and forbidden regions and matched these functions at turning points finding the corresponding connection formulas.

As usual, such a matching proceeds through solutions of the equation with linearised potential in vicinity of turning
points which are given in the case under consideration by the fourth-order Airy functions. Since solutions for the classical momentum $p$
in the classically allowed region $E>V(x)$ in the case of the quartic dispersion contain besides purely real solutions also
purely imaginary solutions, the corresponding semiclassical wave function $\psi(x) \sim \exp(\frac{i}{\hbar}\int^x p(u)du)$ necessarily
contains exponentially increasing and decreasing components. This is in contrast to the conventional case of the quadratic energy-momentum
dispersion where the wave function contains only purely oscillating components in the classically allowed region. Therefore, the correct
account of these exponentially increasing and decreasing components requires determining hyperasymptotics of the fourth-order Airy functions
in the classically allowed region. This conclusion remains true when applying the WKB method for quasiparticles with energy-momentum
dispersion of higher than the fourth power of momentum.

We found the corresponding hyperasymptotics by using the method of steepest descents as well as relating the fourth-order Airy functions to Mainardi's
and Fax$\rm\acute{e}$n's functions (see, the corresponding analysis in Appendix A which provides an important check of our results obtained
via the use of method of steepest descent). Utilizing these hyperasymptotics and matching the semiclassical wave functions in the clssically allowed and
forbidden regions results in four equations whose solution defines a generalized Bohr-Sommerfeld quantization condition with ''instanton-like''
(exponentially small in $1/\hbar$) correction. We emphasize that the hyperasymptotics of fourth-order Airy functions are crucial
to find this non-perturbative correction to the Bohr-Sommerfeld quantization condition which occurs
even for the case of the harmonic potential and quartic dispersion where tunneling and complex turning points are absent. It would be interesting to verify the appearance of such
contributions in other approaches, such as the complex WKB method \cite{Voros}.

Applying the obtained quantization condition to systems with harmonic and quartic potentials, we found that since hyperasymptotics are exponentially
suppressed at large energy, the modification of the quantization condition due to hyperasymptotics is the most relevant for the lowest
energy states. For example, the correction to the WKB perturbative solution due to hyperasymptotics for the WKB ground state energy is around 11$\%$ in the case of quasiparticles with quartic dispersion
and harmonic potential. This correction brings the ground state energy closer to the exact value. It is worth mentioning that the system with quartic dispersion and harmonic potential is
related through the Fourier transform to the Schrödinger equation with anharmonic quartic potential. In this case, our quantization
condition is shown to be in agreement with that obtained in Refs.\cite{Balian,Parisi} where additional turning points in the complex plane were taken
into account.

We would like to add that our findings are directly applicable to the study of bound states in bilayer graphene with for potentials which depend only
on one coordinate. In fact, applying a semiclassical method to studying Zener tunneling in biased bilayer graphene with barrier which depends on one
coordinate, it was revealed in a recent work \cite{Sushkov} the necessity of retaining decaying components even in classically allowed regions in order
to get the correct expressions for the transmission probability. This result agrees with our conclusion that the semiclassical quantization rule for bound
states in the case of the fourth-order operators requires retaining exponentially increasing and decreasing WKB solutions in the classically allowed region
in addition to conventional oscillating solutions. As a result, the quantization condition for bound state energies contains nonperturbative in the Planck
constant correction.

The extension of our results to higher than quartic dispersion $E\sim p^{2n}$ with $n\ge3$ seems straightforward, but requires finding the hyperasymptotics
of higher-order Airy functions satisfying the differential equation of order $2n$.

Finally, our approach can be extended to multi-component systems with linear or quadratic in momentum matrix Hamiltonians such as bilayer graphene
with four or two low-energy bands. In addition, our analysis could be applied to higher spatial dimension systems which admit a separation of variables,
for example, the case of central potentials. Then the system can be reduced to an ordinary higher order differential equation for one of its  components.
From this point of view, it is interesting to extend a recent study of charged impurity induced bound states in bilayer graphene with the screened Coulomb
potential \cite{Shklovskii} and apply the WKB approach developed in this paper.
\vspace{3mm}

\begin{acknowledgments}
\vspace{2mm}
The authors acknowledge support from the National Research Foundation of Ukraine grant
(2023.03/0097) “Electronic and transport properties of Dirac materials and Josephson junctions”.
\end{acknowledgments}
\vspace{8mm}

\appendix

\section{Power series and integral representations for the fourth-order Airy functions $Ai_4(x)$ and
$\widetilde{Ai}_{4}(x)$}
\label{appendix}
\vspace{5mm}

In this appendix we provide some additional useful information on the fourth-order Airy functions $Ai_4(x)$ and $\widetilde{Ai}_{4}(x)$ related to their power series, integral representations, and connection with special functions such as the Wright, Mainardi, and Fax$\rm\acute{e}$n functions.

Using the contour representation (\ref{Ai4-integral}), we obtain a different form for the function $Ai_4(x)$
\begin{align}
Ai_4(x)&= \frac{1}{2\pi i}\int\limits_{e^{-\frac{2\pi}{5}i}\infty}^{e^{\frac{2\pi}{5}i}\infty}e^{-x t-\frac{t^5}{5}}dt
=\frac{1}{2\pi i}\left(\int\limits_0^{e^{\frac{2\pi}{5}i}\infty}-\int\limits_0^{e^{-\frac{2\pi}{5}i}\infty}\right)e^{-x t-\frac{t^5}{5}}dt
\nonumber\\
&=\frac{1}{2\pi i}\left(e^{\frac{2\pi}{5}i}\int\limits_0^\infty\,e^{-xe^{\frac{2\pi}{5}i} t -\frac{t^5}{5}}dt - e^{-\frac{2\pi}{5}i}\int\limits_0^\infty\,e^{-x e^{-\frac{2\pi}{5}i} t -\frac{t^5}{5}}dt\right).
\end{align}
Expanding the integrals in $x$ and integrating over $t$ we get the power series
\begin{align}
Ai_4(x)&=\frac{5^{-4/5}}{\pi}\sum\limits_{n=0}^\infty\frac{(-5^{1/5}x)^n}{n!}\Gamma\bigl(\frac{n+1}{5}\bigr)\cos\frac{(1-4n)\pi}{10}
\nonumber\\
&=\frac{5^{-4/5}}{\pi}\sum\limits_{n=0}^\infty\frac{(-5^{1/5}x)^n}{n!}\Gamma\bigl(\frac{n+1}{5}\bigr)\sin\frac{2(n+1)\pi}{5}
\nonumber\\
&=2\cdot{5^{-4/5}}\sum\limits_{n=0}^\infty\frac{(-5^{1/5}x)^n}{n!\Gamma\bigl(\frac{4-n}{5}\bigr)}\cos\frac{(n+1)\pi}{5},
\label{Ai4-power_series}
\end{align}
which is absolutely converging for all complex $x$. In the last equality, we used the reflection formula for the gamma
function. Similarly, we obtain the following power series representation for $\tilde{Ai}_{4}(x)$:
\begin{align}
\widetilde{Ai}_4(x)&=\frac{5^{-4/5}}{\pi}\sum_{n=0}^\infty\frac{\left(-5^{1/5}x\right)^n}{n!}\Gamma\left(\frac{n+1}{5}\right)\left[1-
\cos\frac{2(n+1)\pi}{5}\right]\nonumber\\
&=\frac{2\cdot 5^{-4/5}}{\pi}\sum_{n=0}^\infty\frac{\left(-5^{1/5}x\right)^n}{n!}\Gamma\left(\frac{n+1}{5}\right)
\sin^2\frac{(n+1)\pi}{5}\nonumber\\
&={2\cdot 5^{-4/5}}\sum_{n=0}^\infty\frac{\left(-5^{1/5}x\right)^n}{n!\Gamma\left(\frac{4-n}{5}\right)}
\sin\frac{(n+1)\pi}{5}.
\label{Ai4-tilde-series}
\end{align}
Using these power series representations we can express the functions $Ai_4(x)$ and $\tilde{Ai}_{4}(x)$ in terms of known special functions.
The Wright function, also known as generalized Bessel function, has the following series representation \cite{Paris2021}:
\begin{align}
W(\lambda,\mu;z)=\sum\limits_{n=0}^\infty\frac{z^n}{n!\Gamma(\lambda n+\mu)}=\frac{1}{\pi}\sum\limits_{n=0}^\infty\frac{z^n}{n!}
\Gamma(1-\mu-\lambda n)\sin\pi(\lambda n+\mu),\qquad \lambda>-1,
\label{Wright-series}
\end{align}
where $\mu$ is complex number.
The function with $-1 < \lambda < 0$ has been termed the Wright function of the second kind and the function with $\lambda > 0$ is referred to
a Wright function of the first kind \cite{Mainardi-book}. It has integral representation in terms of the Mellin-Barnes integral
\begin{align}
W(\lambda,\mu;z)=\frac{1}{2\pi i}\int_{C}(-z)^{-s}\frac{\Gamma(s)}{\Gamma(\mu-\lambda s)}ds,
\end{align}
where the integration contour (the Hankel contour) begins at $s=-\infty$ below the negative axis, encircles $s = 0$ anticlockwise,
and returns to $s=-\infty$ above the negative axis. The second possibility is a contour parallel to the imaginary axis avoiding the
poles of $\Gamma(s)$. In the first case, the integral is convergent for all complex $z$, while in the second case the convergence of the integral
requires $|{\rm arg}(-z)|<\pi(1+\lambda)/2$. Obviously, applying the Cauchy theorem and calculating residues at the poles of $\Gamma(s)$ we come
back at the series representation (\ref{Wright-series}).

The Mainardi function \cite{Mainardi1994,Paris2021} is related to the Wright function of the second kind
\begin{align}
M_\sigma(z)=W(-\sigma,1-\sigma;-z),\qquad 0<\sigma<1.
\label{def:Mainardi}
\end{align}
In terms of Mainardi's function we can write our functions for real $x$ as
\begin{align}
&Ai_4(x)=2\cdot 5^{-4/5}{\rm Re}\left[e^{\pi i/5}M_{1/5}\left(5^{1/5}x\,e^{\pi i/5}\right)\right],\nonumber\\
&\widetilde{Ai}_4(x)=2\cdot 5^{-4/5}{\rm Im}\left[e^{\pi i/5}M_{1/5}\left(5^{1/5}x\,e^{\pi i/5}\right)\right].
\label{Ai4,tildeAi4-vs-Mainardi}
\end{align}
Thus, asymptotics of $Ai_4(x)$ and $\widetilde{Ai}_4(x)$ for $x\to\pm\infty$ are determined by asymptotics of the Mainardi function $M_{1/5}(z)$
when $|z|\to\infty$. The asymptotic behavior of the Wright function of the second kind, and hence of the Mainardi function, has been
carefully studied in Ref.\cite{Wong1999} including the contribution due to hyperasymptotics.

The fourth-order Airy functions under consideration can be expressed in terms of the so-called Fax$\rm{\acute{e}}$n function which is defined by the integral
\begin{align}
{\rm Fi}(a,b;z)=\int\limits_0^\infty dt\,t^{b-1}e^{-t+z t^a}=\sum\limits_{n=0}^\infty\frac{\Gamma(a n+b)}{n!}z^n,\qquad |z|<\infty,
\end{align}
where the parameters satisfy $0<a<1$, $b>0$ and the series is absolutely and uniformly convergent (see, Eq.(5.5.1) in \cite{Paris_MB-book}).

The Mainardi function $M_{\sigma}(z)$ defined in Eq.(\ref{def:Mainardi}) can be expressed in terms of the Fax$\rm\acute{e}n$ function as follows:
\begin{align}
M_{\sigma}(z)=\frac{1}{2\pi}\left[e^{\pi i\theta}{\rm Fi}\left(\sigma,\sigma;z\,e^{-\pi i\kappa}\right)+
e^{-\pi i\theta}{\rm Fi}\left(\sigma,\sigma;z\,e^{\pi i\kappa}\right)\right],\quad \theta=\sigma-\frac{1}{2},\,\, \kappa=1-\sigma,
\label{Mainardi-through-Faxen}
\end{align}
and in a similar manner
\begin{align}
M_{\sigma}(-z)=\frac{1}{2\pi}\left[e^{\pi i\theta}{\rm Fi}\left(\sigma,\sigma;z\,e^{\pi i\sigma}\right)+
e^{-\pi i\theta}{\rm Fi}\left(\sigma,\sigma;z\,e^{-\pi i\sigma}\right)\right].
\end{align}
Combining Eqs.(\ref{Ai4,tildeAi4-vs-Mainardi}) and (\ref{Mainardi-through-Faxen}), we can  write down the fourth-order Airy functions directly in terms of
 Fax\(\acute{e}\)n's function
\begin{align}
\label{Ai4-vs-Faxen}
Ai_4(x)&=\frac{5^{-4/5}}{2\pi i}\left[e^{2\pi i/5}{\rm Fi}\left(-5^{1/5}x\,e^{2\pi i/5}\right)-
e^{-2\pi i/5}{\rm Fi}\left(-5^{1/5}x\,e^{-2\pi i/5}\right)\right],\\
\widetilde{Ai}_4(x)&=\frac{5^{-4/5}}{\pi}\left[{\rm Fi}\left(-5^{1/5}x\right)-\frac{1}{2}e^{2\pi i/5}
{\rm Fi}\left(-5^{1/5}x\,e^{2\pi i/5}\right)-\frac{1}{2}e^{-2\pi i/5}{\rm Fi}\left(-5^{1/5}x\,e^{-2\pi i/5}\right)\right],
\label{Ai4tilde-vs-Faxen}
\end{align}
where we introduced the shorthand notation
\begin{align*}
{\rm Fi}\left(z\right)\equiv {\rm Fi}\left(\frac{1}{5},\frac{1}{5};z\right).
\end{align*}
An important role is played by the connection formula
\begin{align}
{\rm Fi}(a,b;z)=e^{\pm2\pi i b}{\rm Fi}(a,b;z\,e^{\pm2\pi i a})+E_{\pm}(z),
\label{Faxen-identity}
\end{align}
where
\begin{align}
E_{\pm}(z)=\frac{e^{\pm\pi i(b-1/2)}}{2\pi i}\int_{C}\frac{2\pi\Gamma(s)}{\Gamma(1-b+a s)}\left(z\,e^{\mp\pi i(1-a)}\right)^{-s}ds
\label{Faxen-E(z)}
\end{align}
is valid for all ${\rm arg}\,z$ and $0<a<1$. The function $E_{\pm}(z)$ describes an exponential expansion as $|z|\to\infty$ because
the integrand does not contain poles for ${\rm Re}\,z>0$. The connection formula (\ref{Faxen-identity}) relates values of the function
in different sectors. The repeated application of (\ref{Faxen-identity}) yields for
$k = 1, 2,\dots$
\begin{align}
{\rm Fi}(a,b;z)=e^{\pm2\pi i kb}{\rm Fi}(a,b;z\,e^{\pm2\pi ika})+\sum\limits_{r=1}^{k-1}e^{\pm2\pi irb}E_{\pm}(z\,e^{\pm2\pi ira}).
\label{Faxen-identity-repeat}
\end{align}
A comprehensive study of the Fax$\acute{e}$n function, including asymptotics when $|z|\to\infty$, is presented in book \cite{Paris_MB-book}.
Using these results one can derive the asymptotics of the fourth-order Airy functions, including hyperasymptotic terms,
determined by the extended method of steepest descents.

\end{document}